\documentclass[prd,preprint,amsmath,nofootinbib,superscriptaddress]{revtex4}
\usepackage{graphicx}
\usepackage{bm}
\usepackage{epsfig}
\usepackage{color}
\usepackage{colordvi}

\newcommand{\beq}{\begin{equation}}
\newcommand{\eeq}{\end{equation}}
\newcommand{\bea}{\begin{eqnarray}}
\newcommand{\eea}{\end{eqnarray}}
\begin{document}


\title{Asymptotics of $d$-dimensional Kaluza-Klein black holes:  beyond the Newtonian approximation
}

\author{Yi-Zen Chu}
\author{Walter D. Goldberger}
\affiliation{Department of Physics, Yale University, New Haven, CT 06520}
\author{Ira Z.~Rothstein}
\affiliation{Department of Physics, Carnegie Mellon University,
    Pittsburgh, PA 15213  \vspace{0.3cm}}
\begin{abstract}

We study the thermodynamics of small black holes in compactified spacetimes of the form ${\bf R}^{d-1}\times {\bf S}^1$.  This system is analyzed with the aid of an effective field theory (EFT) formalism in which the structure of the black hole is encoded in the coefficients of operators in an effective worldline Lagrangian.   In this effective theory, there is a small parameter $\lambda$ that characterizes the corrections to the thermodynamics due to both the non-linear nature of the gravitational action as well as effects arising from the finite size of the black hole.   Using the power counting of the EFT we show that the series expansion for the thermodynamic variables contains terms that are analytic in $\lambda$,  as well as certain fractional powers that can be attributed to finite size operators.  In particular our operator analysis shows that existing analytical results do not probe effects coming from horizon deformation.  As an example, we work out the order $\lambda^2$ corrections to the thermodynamics of small black holes for arbitrary $d$, generalizing the results in the literature.   

 \end{abstract}

\maketitle

\section{Introduction}

General relativity in spacetime dimension $d$ larger than four supports black brane solutions that, unlike in lower dimensions, are not uniquely characterized by their asymptotic charges (mass, spin, gauge charges).   An example of this situation is the Kaluza-Klein black hole, a solution of the Einstein equations consisting of a black hole embedded in a compactified spacetime, for instance ${\mathbf R}^{d-1}\times {\mathbf S}^1$.    Because of the lack of uniqueness in $d\geq 4$, this system exhibits a range of phases, characterized by the horizon topology, as the period $L$ of the ${\bf S}^1$ is varied.   For $L$ much larger than the horizon length scale, the horizon topology is ${\bf S}^{d-2}$ corresponding to an isolated black hole.   As $L$ becomes of order $r_s$ one finds uniform and non-uniform black string phases with horizon topology ${\bf S}^{d-3}\times {\bf S}^1$.   There is evidence to support the conjecture that  uniform string decays~\cite{gl1,gl2} proceed via a topology changing phase transition into a black hole final state (see~\cite{rev1,rev2} for reviews).  Other proposals for the final state of the unstable black string can be found in~\cite{o1,o2}.


Understanding the dynamics of the black hole/black string phase transition is important for a variety of reasons.   Apart from being a toy model for studying the physics of topology change in higher dimensional general relativity, it is also relevant for its connection to gauge/gravity duality in string theory~\cite{gg1,gg2}.   Also, the Kaluza-Klein black hole plays a role in the phenomenology of scenarios where gravity is strong at the TeV scale, and production of higher dimensional black holes at the LHC becomes a possibility.

There does not exist an analytic solution of the Einstein equations describing a black hole in the background  ${\bf R}^{d-1}\times {\bf S}^1$ with $d\geq 5$ (however, see~\cite{ho1}; for $d=4$, a closed form metric can be found in ref.~\cite{myers}).  For generic values of the ratio $r_s/L$ one must resort to numerical simulations in order to find solutions.    These have been carried out in~\cite{kolnumeric,kudoh1,kudoh2}.   Here, we will consider the asymptotic region of the phase diagram in which the parameter $\lambda=(r_s/L)^{d-3}$ is much less than unity, and analytic solutions can be found perturbatively.  Although this region of parameter space is likely to be far from where the black hole/black string transition is expected to take place, it is a region that can be mapped out analytically.  These perturbative calculations provide a useful test of the numerical simulations, and by extrapolation, may give qualitative information on the full phase diagram of solutions.

The $\lambda$ corrections to the thermodynamics of a small black hole in the background ${\mathbf R}^{d-1}\times {\mathbf S}^1$ have been calculated in ref.~\cite{hd,kold} to leading order for arbitrary $d$, and in ref.~\cite{5d} to order $\lambda^2$ for $d=5$.   In ref.~\cite{hd}, the order $\lambda$ corrections were calculated by employing a specialized coordinate system~\cite{ho1} for the entire spacetime.  Alternatively, the approach taken in~\cite{kold,5d} is to split the spacetime into a region near the black hole where the solution is the $d$-Schwarzschild metric,
\begin{eqnarray}
\label{eq:BHmetric}
\nonumber
ds^2 &=& f(r) dt^2 - {1\over f(r)} dr^2- r^2d\Omega_{d-2}^2,\\
f(r) &=& 1- \left({r_s\over r}\right)^{d-3},
\end{eqnarray}
 weakly perturbed by compactification, and a far region in which the metric can be parametrized in terms of asymptotic multipole moments  (see ref.~\cite{kolPT} for a systematic discussion of this procedure).   These two solutions are then patched together in an overlap region, yielding a relation between the short distance parameters (the scale $r_s$ of the $d$-dimensional Schwarzschild metric) and the mass $m$ and tension $\tau$ as measured by an observer far from the black hole\footnote{Note, however, that ultraviolet divergences arise in the computation of the asymptotic metric coefficients already at leading order in $\lambda$.   This behavior can be traced to the short distance singularities of the $d$-dimensional flat space Green's function.  A prescription for  handling such divergences at leading order in $\lambda$ can be found in~\cite{kold}.}.   As discussed in~\cite{ho2,koltheory}, all thermodynamic quantities relevant to the phase diagram can be calculated given the asymptotic ``charges" $m,\tau$.   

Here, we propose a different method for calculating the phase diagram in the perturbative region $\lambda\ll 1$, based on the effective field theory approach applied  to extended gravitational systems developed in~\cite{GnR1,GnR2}.    Since in the $\lambda\ll 1$ limit there is a large hierarchy between the short distance scale $r_s$ and the compactification size, it is natural to integrate out ultraviolet modes at distances shorter than $r_s$ to obtain an effective Lagrangian describing the dynamics of the relevant degrees of freedom at the scale $L$.    In the resulting EFT, the scale $r_s$ only appears in the Wilson coefficients of operators in the action constructed from the relevant modes.  Ignoring horizon absorption~\cite{GnR2} and spin~\cite{porto}, these long wavelength modes are simply the metric tensor $g_{\alpha\beta}$ coupled to the black hole worldline coordinate $x^\mu(s)$.  The couplings of the particle worldline to the metric can be obtained by a fairly straightforward matching calculation, although one expects that all operators consistent with symmetries (diffeomorphism invariance, worldline reparametrizations) are present.  

Although clearly there are some similarities between the EFT approach and the matched asymptotics of~\cite{kold,5d,kolPT}, there are several advantages to formulating the $\lambda$ expansion in the language of an EFT:
\begin{itemize}
\item  In the EFT, it is possible to disentangle the terms in the perturbative expansion that arise from the finite extent of the black hole, which scale like integer powers of $r_s/L$, versus ``post-Newtonian" corrections due to the  non-linear terms in the Einstein-Hilbert Lagrangian that scale like integer powers of 
\begin{equation}
v^2 \equiv {2 G_N m_0\over L^{d-3}} = \left({r_s\over L}\right)^{d-3},
\end{equation}
and are therefore also equivalent to powers of $\lambda$.   
\item The EFT has manifest power counting in $\lambda$.   This means that it is possible to determine at what order in the expansion effects from the finite size of the black hole horizon first arise.   As we will show in the next section, the first finite size correction, which in the EFT manifests itself through a non-minimal coupling of the black hole worldline to the Riemann tensor arises  at order $\lambda^{(d+1)/(d-3)}=(r_s/L)^{d+1}$.  For a fixed $d$ the finite size effects, for example the tidal distortion of the black hole horizon, will contribute to the thermodynamic variables at order $\lambda^{\frac{2d-2}{d-3}}$ relative to the leading order result.  Thus the finite horizon effects become as large as $\lambda^2$ as $d\rightarrow\infty$.    
This also indicates that the results of refs.~\cite{kold,5d} are not sensitive to the specific structure of the Kaluza-Klein black hole, but rather reflect the thermodynamics of structureless point particles.

\item In the EFT, calculations can be carried out using the standard tools of field theoretic perturbation theory.   In particular, the perturbative expansion has a diagrammatic interpretation in terms of standard Feynman diagrams.   Ultraviolet divergences that arise in Feynman integrals can be dealt with using a standard regulator (e.g, dimensional regularization) and absorbed into the coefficients of local operators.   There is no impediment to renormalizing the theory to all orders in $\lambda$.   As an example of this procedure we calculate in sec.~\ref{sec:PT}  the ${\cal O}(\lambda^2)$ corrections to the asymptotic mass and tension of the Kaluza-Klein black hole.
\end{itemize}

Our results are organized as follows.  In sec.~\ref{sec:EFT} we formulate the EFT and derive the power counting rules for $\lambda\ll 1$.   Using this power counting we analyze the relative contribution of an arbitrary finite size worldline operator.  In sec.~\ref{sec:PT} we use the EFT to calculate the ${\cal O}(\lambda^2)$ corrections to the asymptotic charges $m$ and $\tau$ for arbitrary $d$ and use these results in sec.~\ref{sec:thermo} to work out the corresponding corrections to the thermodynamic relations.   In this section we also compare our analytic formulas to the results of numerical simulations~\cite{kudoh1,kudoh2} for $d=5$ and $d=6$.

\section{The Effective Field Theory}
\label{sec:EFT}

We consider an isolated black hole in a background spacetime of the form ${\bf R}^{d-1}\times {\bf S}^1$.   Coordinates on ${\bf R}^{d-1}$ are denoted by $x^m=(x^0,{\bf x})$ and $z$ labels circumference along ${\bf S}^1$.   Coordinates on $d$-dimensional spacetime are denoted $x^\mu=(x^m,z)$.      The period of the ${\bf S}^1$ factor as measured by an observer at $r=|{\bf x}|\rightarrow\infty$ is $L$.

In order to determine the phase diagram of this system, it  is sufficient to calculate the moments of the Kaluza-Klein black hole that appear in the first non-trivial corrections to the asymptotic metric.   By the symmetries of the background, the non-vanishing terms are, to leading order as $r\rightarrow\infty$,
\begin{eqnarray}
g_{00} &=& 1 +{c_{00}\over r^{d-4}}, \\
 g_{zz}  &=& -1 +{c_{zz}\over r^{d-4}}.
\end{eqnarray}
The coefficients $c_{00}$ and $c_{zz}$ are related to the asymptotic mass $m$ and tension $\tau$ by the relations~\cite{ho2,koltheory},
\begin{eqnarray}
\label{eq:ab}
\left(\begin{array}{c}
c_{00}\\
c_{zz}
\end{array}\right)= {4\pi^{1/2} G_N\over L}{\Gamma(d/2-2)\over \Gamma(d/2-1/2)} \left(
\begin{array}{cc}
-(d-3)  & 1 \\
- 1&  (d-3)
 \end{array}\right) 
 \left(\begin{array}{c}
 m \\
 \tau L
 \end{array}\right)
\end{eqnarray}  
The constant $G_N$ is defined such that the Newton potential between two masses in uncompactified $d$-dimensional space is $V(r)=-G_N m_1 m_2/r^{d-3}$.

In the limit $\lambda = (r_s/L)^{d-3}\ll 1$, these quantities can be calculated in perturbation theory.   One method is to solve the Einstein equations perturbatively, using the matched asymptotic techniques of~\cite{kold,5d,kolPT}.   Another possibility is to first integrate out the black hole, replacing the spacetime in the vicinity of the horizon with an effective Lagrangian for the black hole worldline coupled to gravity.   Including all  terms with up to two derivatives (we will be more specific about the expansion parameter in this expansion below) this Lagrangian takes the form
\begin{equation}
\label{eq:EFTS}
S=- 2 m^{d-2}_p\int d^d x \sqrt{g} R[g] -m_0\int ds +c_E\int ds  E_{\alpha\beta} E^{\alpha\beta} + c_B\int ds B_{\alpha_1\cdots\alpha_{d-2}} B^{\alpha_1\cdots\alpha_{d-2}} +\cdots.
\end{equation}
Here, as in any other EFT,  we have simply written down all terms compatible with diffeomorphism invariance and worldline reparametrization invariance.   In this equation, $ds^2 = g_{\alpha\beta} dx^\alpha dx^\beta$ and the tensors $E_{\alpha\beta}$, and $B_{\alpha_1\cdots\alpha_{d-2}}$ are the electric and magnetic components of the Riemann tensor.   
\begin{eqnarray}
E_{\alpha\beta} &=& R_{\alpha\gamma\beta\delta} {d x^\gamma\over ds} {d x^\delta\over ds},\\
B_{\alpha_1\cdots\alpha_{d-2}} &=& {1\over (d-2)!} {dx^\sigma\over ds} {dx^\rho\over ds} \epsilon_{\sigma\alpha_1\cdots\alpha_{d-3}\mu\nu} {R^{\mu\nu}}_{\rho\alpha_{d-2}}.
\end{eqnarray}
Note that if the black hole is in a Ricci flat background then operators involving the Ricci tensor can be removed by field redefinitions of the metric.   In this case the components of $E_{\alpha\beta}, B_{\alpha_1\cdots\alpha_{d-2}}$ are sufficient to specify the Riemann tensor.   All coefficients of operators in Eq.~(\ref{eq:EFTS}) scale like powers of $m_p$ and  $r_s$, given by
\begin{equation}
r^{d-3}_s  =  2 G_N m_0 =  2^{d-4}{m_0\over m^{d-2}_p}  {d-3\over d-2} {\Gamma((d-3)/2)\over (4\pi)^{(d-1)/2}}
\end{equation}
in a way that we can be fixed by matching to the full Schwarzschild solution (see below).

Starting from Eq.~(\ref{eq:EFTS}),  we use the background field method~\cite{DeWitt,V} to calculate $m$ and $\tau$.  We decompose the metric tensor into a long wavelength non-dynamical background field ${\bar g}_{\alpha\beta}$ and a short wavelength graviton field $h_{\alpha\beta}/m^{d/2-1}_p$
\begin{equation}
g_{\alpha\beta} = {\bar g}_{\alpha\beta} + {h_{\alpha\beta}\over m^{d/2-1}_p},
\end{equation}
and do the path integral over $h_{\alpha\beta}$, holding the black hole worldline to some fixed value $x^\mu(s)$,
\begin{equation}
\label{eq:PI}
\exp i \Gamma_{eff}[{\bar g},x] = \int {\cal D} h_{\alpha\beta}  \exp i \left(S[{\bar g}+h,x] + S_{GF}[{\bar g},h]\right),
\end{equation}
where $S_{GF}$ is a suitable gauge fixing term.  It is convenient to choose $S_{GF}$ to be compatible with background field diffeomorphisms, for example
\begin{equation}
S_{GF} = m^{d-2}_p \int d^d x\sqrt{\bar g}  {\bar g}^{\alpha\beta} \Gamma_\alpha\Gamma_\beta,
\end{equation}
with $\Gamma_\alpha  = {\bar D}^\mu h_{\mu\alpha} - {1\over 2} {\bar D}_\alpha{h^\mu}_\mu$.   

\begin{figure*}[t!]
 \includegraphics[width=8cm]{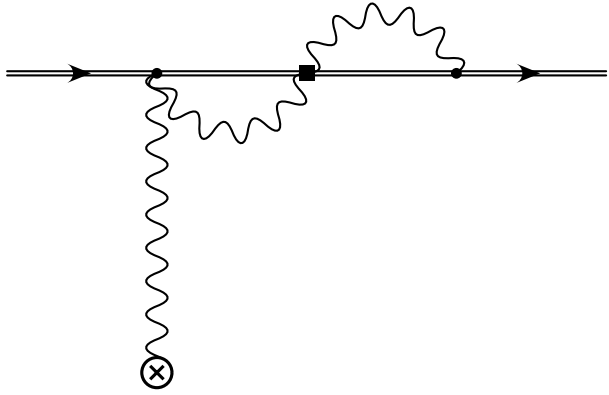} 
 \caption{Leading order contribution of the operators ${\cal O}_{E,B}$ to the effective action tadpoles.   The thick square vertex denotes an insertion of ${\cal O}_{E,B}$, and the $\otimes$ corresponds to an insertion of the background graviton field.}
\label{eb}
\end{figure*}

To calculate Eq.~(\ref{eq:PI}) it is sufficient to linearize about flat space,
\begin{equation}
{\bar g}_{\alpha\beta} = \eta_{\alpha\beta} + {\bar h}_{\alpha\beta},
\end{equation}
and to take $x^\mu=(t,{\bf x}=0,z=0)$.   The relation between $m, \tau$ and $m_0, L$ can be read off the linear terms in $\Gamma_{eff}[{\bar h}]$ with no derivatives
\begin{equation}
\label{eq:tad}
\Gamma_{eff}[{\bar h}] = -{1\over 2} m\int dt {\bar h}_{00} + {1\over 2} \tau L\int dt {\bar h}_{zz} +\cdots.
\end{equation}
Because of Eq.~(\ref{eq:PI}), these two terms are simply the sum of Feynman diagrams like those of Fig.~\ref{eb}, Fig.~\ref{m2} and Fig.~\ref{m3}.  Wavy internal lines denote the propagator for the graviton $h_{\alpha\beta}$, which given our form for $S_{GF}$ is
\begin{equation}
D_{\alpha\beta;\mu\nu}(x-x';z-z') = P_{\alpha\beta;\mu\nu} D(x-x',z-z'),
\end{equation}
with $P_{\alpha\beta;\mu\nu}={1\over 2} \left[\eta_{\alpha\mu}\eta_{\beta\nu} + \eta_{\alpha\nu}\eta_{\beta\mu} - {2\over d-2} \eta_{\alpha\beta}\eta_{\mu\nu}\right]$, and
\begin{equation}
D(x-x';z-z') = {1\over L}\sum^\infty_{n=-\infty} \int {d^{d-1} k\over (2\pi)^{d-1}} {i\over k^2 - (2\pi n/L)^2} e^{-i k\cdot (x-x') +  2 \pi i  n(z-z')/L},
\end{equation}
is the Kaluza-Klein representation of the propagator on flat ${\bf R}^{d-1}\times {\bf S}^1$.   The solid lines denote the black hole worldline.   There are no propagators associated with such lines.  An external line denotes an insertion of a factor of ${\bar h}_{\alpha\beta}$.   Finally, the vertices are constructed from the $n$-graviton terms in the expansion of Eq.~(\ref{eq:EFTS}) about flat space.  Diagrams that  become disconnected by the removal of the particle worldline do not contribute to the terms in $\Gamma_{eff}[{\bar h}]$.  Note also that if we treat Eq.~(\ref{eq:tad}) as an effective source term in the Einstein equations, we recover the relation Eq.~(\ref{eq:ab}) between the metric coefficients $c_{00},c_{zz}$ and the thermodynamic charges.

Each Feynman diagram in the EFT contributes a definite power of $\lambda$ to the terms in Eq.~(\ref{eq:tad}).   Counting powers of $\lambda$ is straightforward.   Given that the only scale in the propagator is $L$ we assign $x^\mu\sim L$ and thus
\begin{equation}
D_{\alpha\beta;\mu\nu} \sim \int {k^{d-1} dk \over k^2} \sim L^{2-d},
\end{equation}   
so that we assign the scaling $h_{\alpha\beta}\sim L^{1-d/2}$.   We assign no power counting factors to ${\bar h}_{\alpha\beta}$.    Power counting relative to the action for the free background graviton, the parameter  $\ell$ that counts graviton loops is
\begin{equation}
m^{d-2}_p\int d^d x {\bar h} \partial^2 {\bar h}\sim (m_p L)^{d-2}\equiv \ell^{-1}.
\end{equation}
This means that 
\begin{equation}
\label{eq:LO}
m_0 \int dt {\bar h}_{00}\sim m_0 L\sim \ell^{-1}\lambda,
\end{equation}
and for example the terms
\begin{eqnarray}
V_1 &=& -{m_0\over 2 m^{d/2-1}_p}\int dt h_{00},\\
V_2 &=&  {m_0\over 4 m^{d/2-1}_p}\int dt h_{00} {\bar h}_{00} 
\end{eqnarray}
obtained by expanding Eq.~(\ref{eq:EFTS}), scale as $V_1, V_2 \sim \ell^{-1/2}\lambda$.   Therefore the diagram in Fig.~\ref{m2}(b) gives a contribution to $\Gamma_{eff}[{\bar h}]$ that scales like ($\langle\cdots\rangle$ denotes a time ordered VEV in the free graviton theory),
\begin{equation}
\mbox{Fig.}~\ref{m2}(b) = \langle  V_1 V_2\rangle \sim \ell^{-1} \lambda^2,
\end{equation}
so that by Eq.~(\ref{eq:LO}) it gives a contribution to $m$ that is suppressed by a single power of $\lambda$ relative to the leading order result $m=m_0,\tau=0$.  Likewise, the ${\bar h} h^2$ vertex in the gravitational action is
\begin{equation}
V_3=\int d^d x {\bar h} \partial h\partial h \sim \ell^{0}\lambda^0
\end{equation}
so that Fig.~\ref{m2}(a) scales like $\langle  V^2_1 V_3\rangle\sim \ell^{-1} \lambda^2$.  In general, a given diagram scales as $\ell^g \lambda^n$, where $n\geq 0$ and $g\geq -1$, the latter bound saturated by diagrams containing no internal graviton loops.

In order to power count the worldline operators with more derivatives, we first need to fix the dependence of the coefficients on $m_p$ and $r_s$.   This is done by matching the effective Lagrangian of Eq.~(\ref{eq:EFTS}) to the full black hole theory, described by the Schwarzschild metric of Eq.~(\ref{eq:BHmetric}).     As in any other EFT, the matching procedure consists of adjusting the couplings in the effective Lagrangian so that observables calculated in the EFT agree with those of the full theory.   

A convenient observable to match to is the $S$-matrix element for low energy elastic graviton scattering off the black hole geometry.   In the full theory, this is obtained by solving the linearized wave equation for the graviton field in the Schwarzschild metric.   After separation of variables this boils down to solving a radial equation that generalizes the Regge-Wheeler equation describing perturbations of four-dimensional Schwarzschild black holes to $d$-dimensions.   The explicit form of this equation can be found in~\cite{RWpt1,RWpt2} (see also~\cite{kolPT}).   Since the only scale in the full theory is $r_s$ we expect the amplitude to take the form\footnote{The norm of the states is irrelevant as it will cancel in the matching.}
\begin{equation}
\langle k',h' | S | k,h\rangle = (2\pi)\delta(k_0'-k_0) r^{d-3}_s f_{hh'}(r_s\omega,\theta)
\end{equation}
where $\omega$ is the energy of the incident graviton, $h,h'$ are spin labels, $\theta$ is the scattering angle and $f_{hh'}$ is a calculable function.   

In the EFT, the scattering cross section receives contributions from insertions of all the couplings in Eq.~(\ref{eq:EFTS}).  In particular, the two-derivative operators give rise to terms in the amplitude that go like
\begin{equation}
\left.\langle k',h' | S | k,h\rangle\right|_{E,B} = (2\pi)\delta(k_0'-k_0){\omega^4\over m^{d-2}_p} \left[c_E E_{hh'}(\theta) + c_B B_{hh'}(\theta)\right],
\end{equation}
where $E_{hh'},B_{hh'}$ are functions whose specific form is not important for our purposes here.  Thus $c_{E,B}$ are non-zero only if $f_{hh'}$ has a term in the low energy limit that scales like $\omega^4$.   If this is the case then we find that $c_{E,B}\sim r^{d+1}_s m^{d-2}_p$.   After expanding about flat space, we have for ${\cal O}_E=c_{E}\int ds E_{\alpha\beta} E^{\alpha\beta}$, ${\cal O}_B=c_{B}\int ds B_{\alpha_1\cdots\alpha_{d-2}} B^{\alpha_1\cdots\alpha_{d-2}}$,
\begin{equation}
{\cal O}_{E,B}\rightarrow r^{d+1}_s \int dt \partial^2 h \partial^2 h\sim \lambda^{d+1\over d-3}.
\end{equation}
The first contribution to the tadpoles in $\Gamma_{eff}[{\bar h}]$ due to an insertion of ${\cal O}_{E,B}$ is from the diagram in Fig.~\ref{eb}.    According to our power counting rules
\begin{equation}
\mbox{Fig.}~\ref{eb} =\langle  V_1 V_2 {\cal O}_{E,B}\rangle\sim \ell^{-1} \lambda^{3d-5\over d-3},
\end{equation}
implying that the $\lambda\ll 1$ thermodynamics is not sensitive to the structure of the black hole until order $\lambda^{2(d-1)/d-3}$, which for $d=5$ is one order beyond the second order results of~\cite{5d} and becomes ${\cal O}(\lambda^2)$ as $d\rightarrow\infty$.   More generally, a worldline operator with $p$ derivatives and $q$ factors of the graviton scales like
\begin{equation}
{c_{p,q}\over m_p^{k(d/2-1)}}\int dt \partial^p h^q\sim \ell^{(q-2)/2}\lambda^{p/(d-3)}
\end{equation}
and it gives a contribution to the charges $m$ and $\tau$ that is order  $\lambda^{q+p/(d-3)}$ ($p\geq4$).

\section{Asymptotic charges}
\label{sec:PT}

As an application of the EFT method, we now compute the ${\cal O}(\lambda^2)$ corrections to the quantities $m, \tau$ that  govern the thermodynamics of the Kaluza-Klein black hole.   According to the power counting rules established in the previous section, the relevant diagrams are those of Fig.~\ref{m2} and Fig.~\ref{m3}.   Finite size effects do not come in at this order.

\subsection{Order $\lambda$}

\begin{figure*}[t!]
\def\size{8 cm}
\hbox{\vbox{\hbox to \size {\hfil \includegraphics[width=7cm]{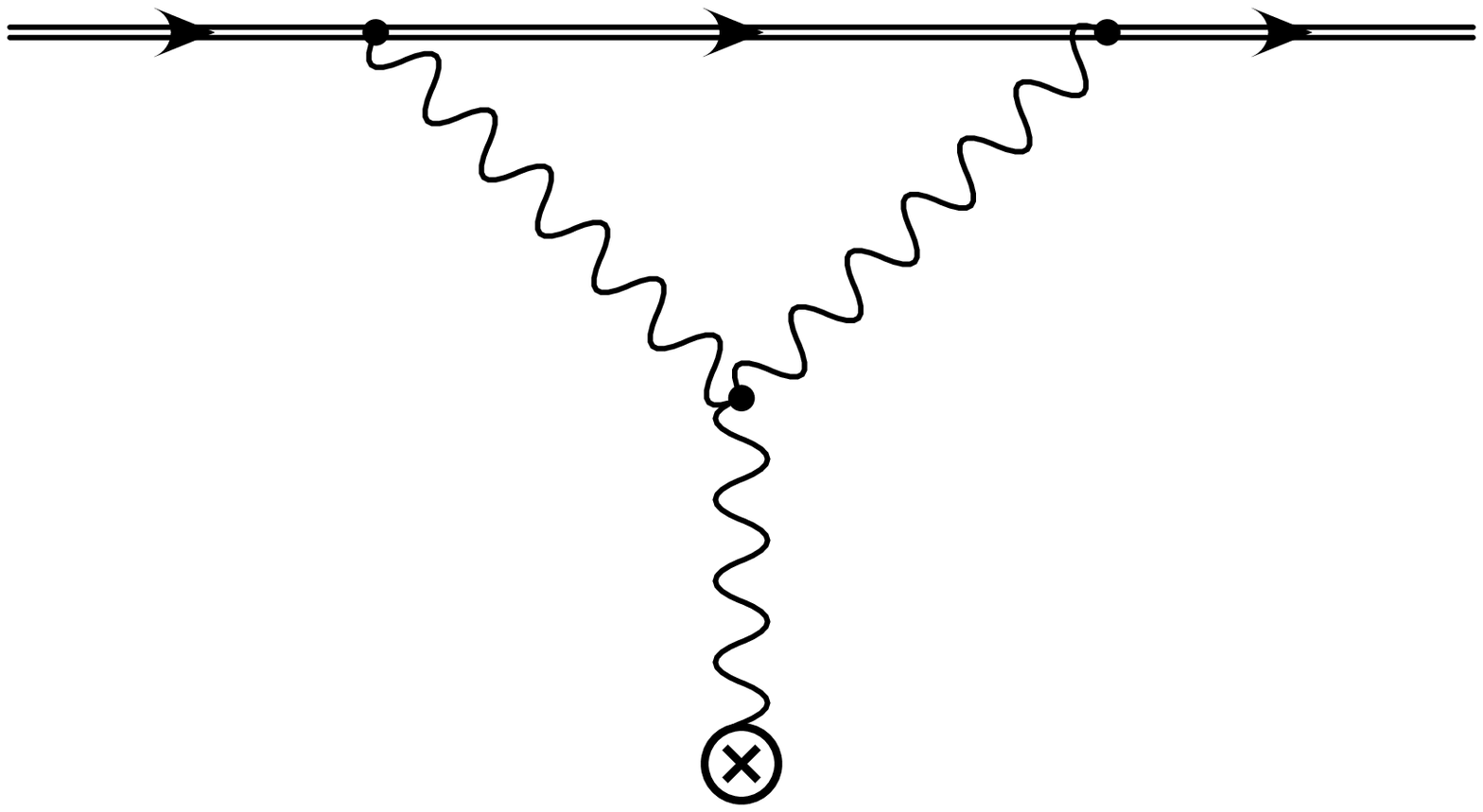} \hfil }\hbox to \size {\hfil(a)\hfil}}
\vbox{\hbox to \size {\hfil \includegraphics[width=7cm]{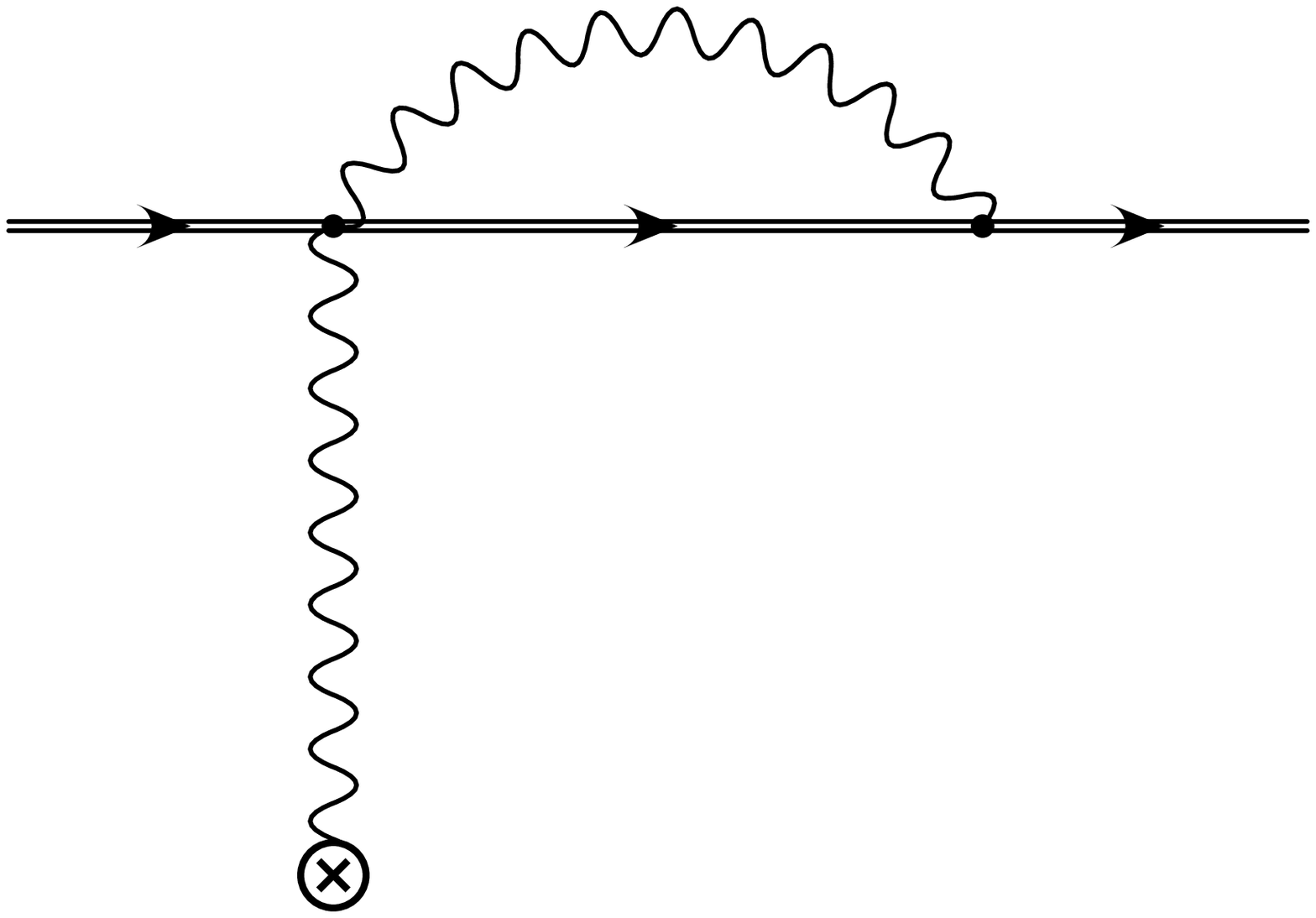} \hfil}\hbox to \size {\hfil(b)\hfil}}
}
\caption{Diagrams contributing to the background field tadpoles at order $\ell^{-1}\lambda^2$.  The $\otimes$ denotes an insertion of the background graviton field.}
\label{m2}
\end{figure*}

The first corrections to the mass and tension of the system arise from the two diagrams in Fig.~\ref{m2}.   The diagram in Fig.~\ref{m2}(a) gives a contribution to the background field effective action which, using the Feynman rules of the EFT, is of the form
\begin{equation}
\label{eq:v}
\mbox{Fig.}~\ref{m2}(a) = {1\over 2!} \left({-i m_0\over 2 m^{d/2-1}_p}\right)^2 \left({-2 i\over m^{d/2-1}_p}\right)\int dt {1\over L}\sum^\infty_{n=-\infty}\int {d^{d-2} k_\perp\over (2\pi)^{d-2}}\left[{1\over k_\perp^2 + (2\pi n/L)^2}\right]^2 V_{k_\perp n}({\bar h}),
\end{equation}
where the vertex function is
\begin{equation}
V_{k_\perp n}({\bar h}) = {d-3\over d-2}\left[ {1\over 2} k_\perp^\mu k_\perp^\nu -(k_\perp^2 + (2\pi n/L)^2)
\left(v^\mu v^\nu -{1\over 4}\eta^{\mu\nu}\right)\right] {\bar h}_{\mu\nu}.
\end{equation}
Since we are only interested in the terms of Eq.~(\ref{eq:tad}) we have set ${\bar h}_{\mu\nu}$ to a constant, in which case the momentum flowing into the diagram vanishes and the calculation of the integral simplifies.  For the ${\bar h}_{00}$ term, Eq.~(\ref{eq:v}) gives
\begin{equation}
\mbox{Fig.}~\ref{m2}(a) = {3i\over 4} m_0 \left({r_s\over L}\right)^{d-3}\zeta(d-3)\int dt {\bar h}_{00},
\end{equation}
where we have used
\begin{equation}
I_0(L)\equiv {1\over 2 L}\sum^\infty_{n=-\infty}\int {d^{d-2} k_\perp\over (2\pi)^{d-2}}{1\over k_\perp^2 + (2\pi n/L)^2} =  {\Gamma\left({d-3\over 2}\right)\over (4\pi)^{d-1\over 2}} \left({2\over L}\right)^{d-3}\zeta(d-3),
\end{equation}
where $\zeta(z)$ is the Riemann zeta function.  Note that this integral is actually ultraviolet divergent.   The divergence renormalizes the point particle mass and can be absorbed by a shift in $m_0$.   We use dimensional regularization to deal with this.  Since we are interested in $d\geq 5$, the divergent part of the integral is simply set to zero by the regulator.    

For the ${\bar h}_{zz}$ term we have
\begin{equation}
\mbox{Fig.}\ref{m2}(a) = {i\over 4} m_0\left({r_s\over L}\right)^{d-3}(d-3)\zeta(d-3)\int dt {\bar h}_{zz}.
\end{equation}
Here we have used the additional integral
\begin{equation}
I_1(L)\equiv {1\over 2L}\sum^\infty_{n=-\infty} \int {d^{d-2} k_\perp\over (2\pi)^{d-2}}\left[{2\pi n/L\over k_\perp^2 + (2\pi n/L)^2}\right]^2 =\left(2-{d\over 2}\right) I_0(L).
\end{equation}

Since the source is at rest, Fig.~\ref{m2}(b) only gives a contribution to the ${\bar h}_{00}$ tadpole
\begin{equation}
\mbox{Fig.}~\ref{m2}(b) = -{i\over 2} m_0 \left({r_s\over L}\right)^{d-3}\zeta(d-3)\int dt {\bar h}_{00}.
\end{equation}
Combining these results we find to first order in $\lambda$
\begin{eqnarray}
{m\over m_0}  &=& 1-{1\over 2}{\hat\lambda}\\
{\tau L \over m_0} &=& {1\over 2} (d-3){\hat\lambda},
\end{eqnarray}
where we have defined ${\hat\lambda}= \zeta(d-3) \left({r_s\over L}\right)^{(d-3)}$.  This reproduces the results of~\cite{hd,kold}.

\subsection{Order $\lambda^2$}

\begin{figure*}[t!]
\def\size{8 cm}
\hbox{\vbox{\hbox to \size {\hfil \includegraphics[width=7cm]{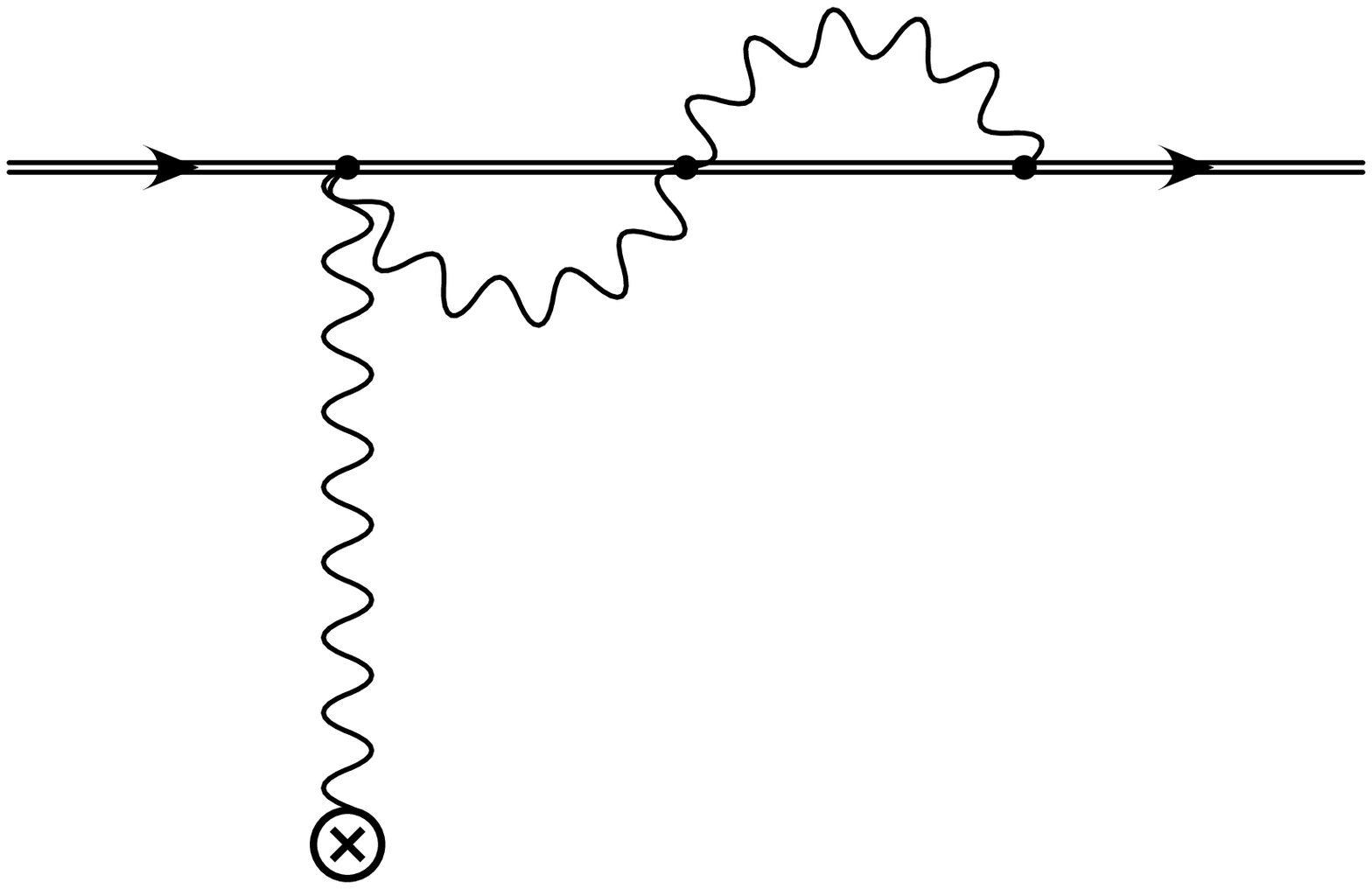} \hfil }\hbox to \size {\hfil(a)\hfil}}
\vbox{\hbox to \size {\hfil \includegraphics[width=7cm]{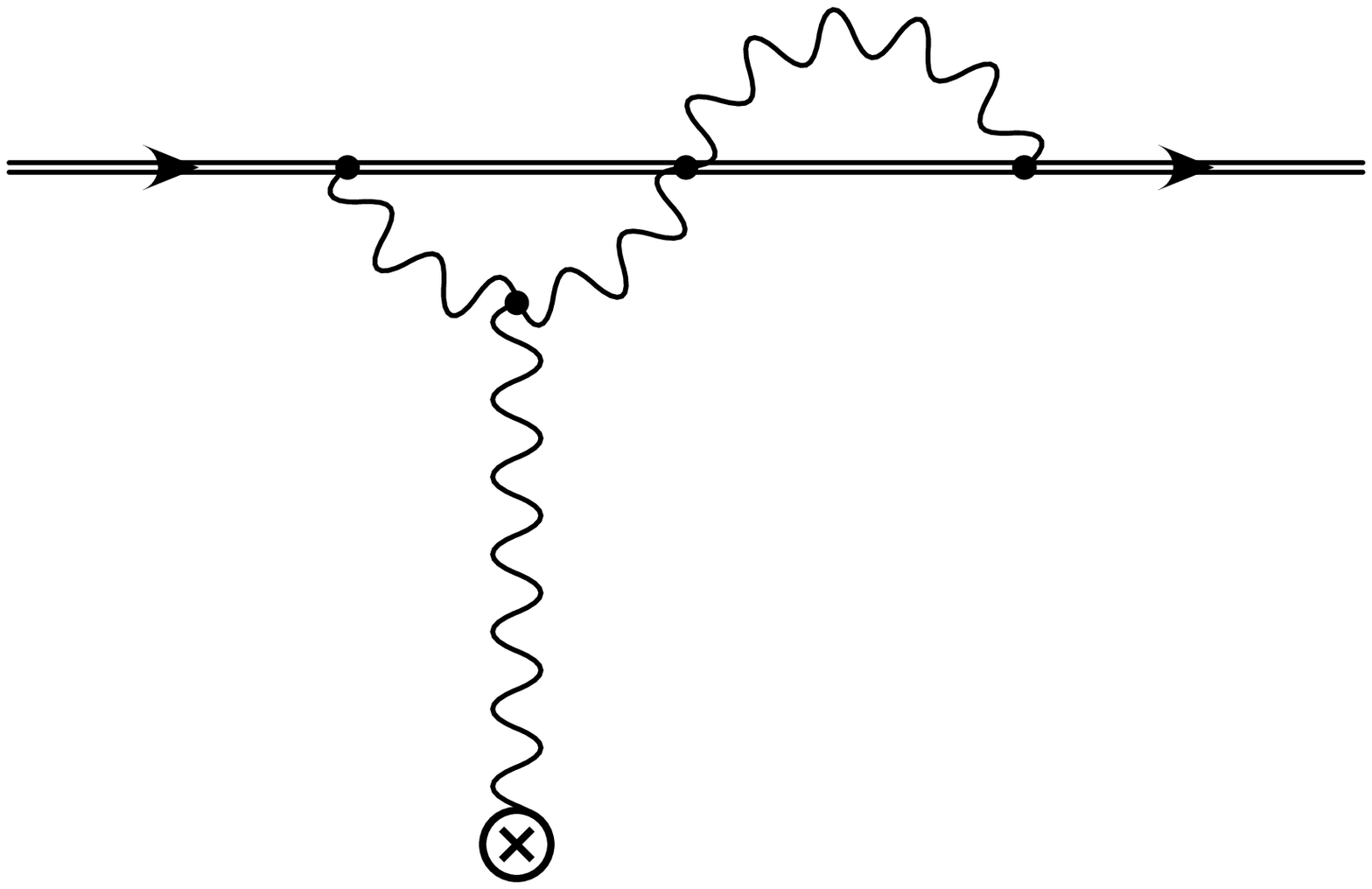} \hfil}\hbox to \size {\hfil(b)\hfil}}}
\vspace{.5cm}
\hbox{\vbox{\hbox to \size {\hfil \includegraphics[width=7cm]{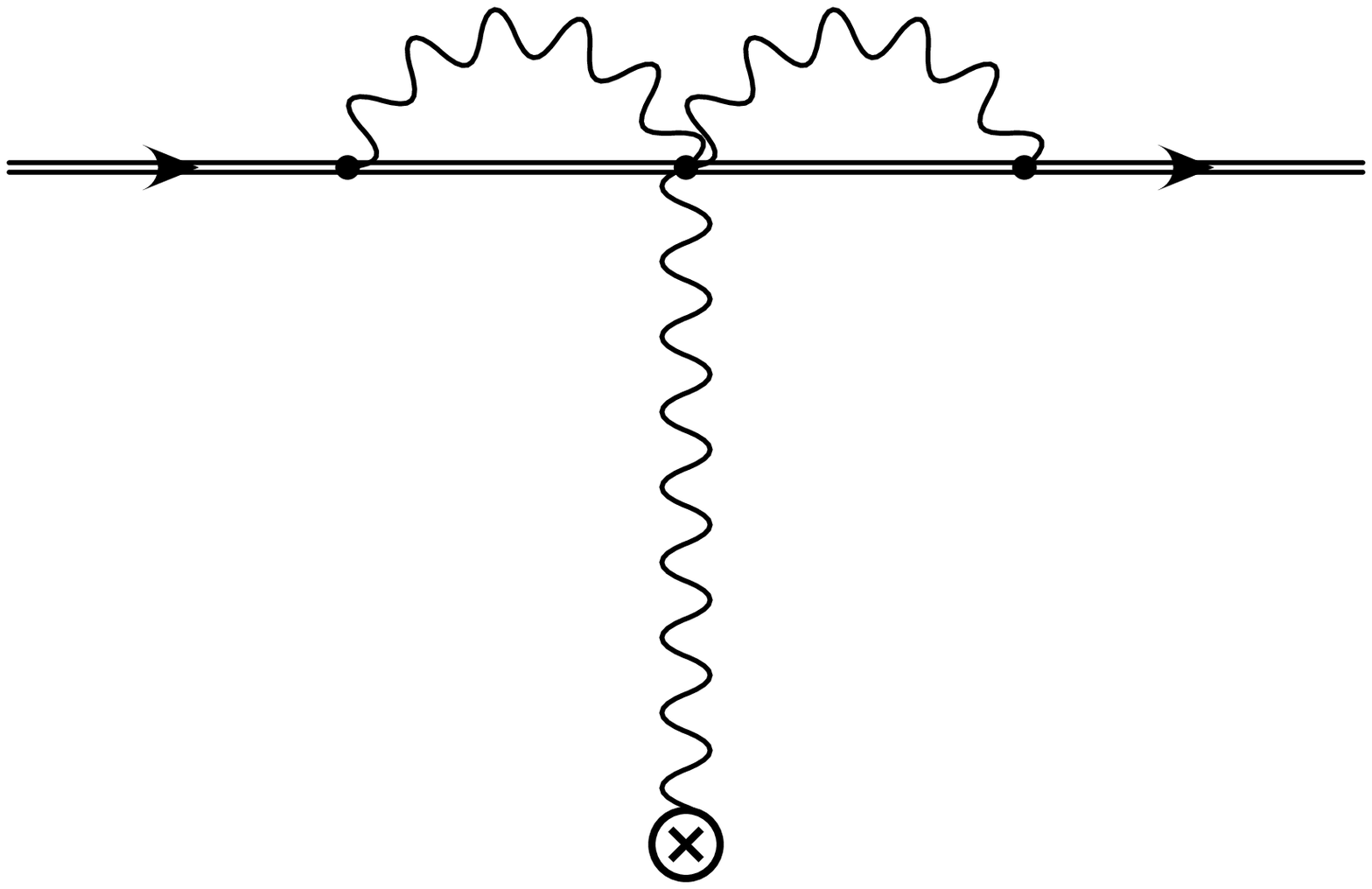} \hfil }\hbox to \size {\hfil(c)\hfil}}
\vbox{\hbox to \size {\hfil \includegraphics[width=7cm]{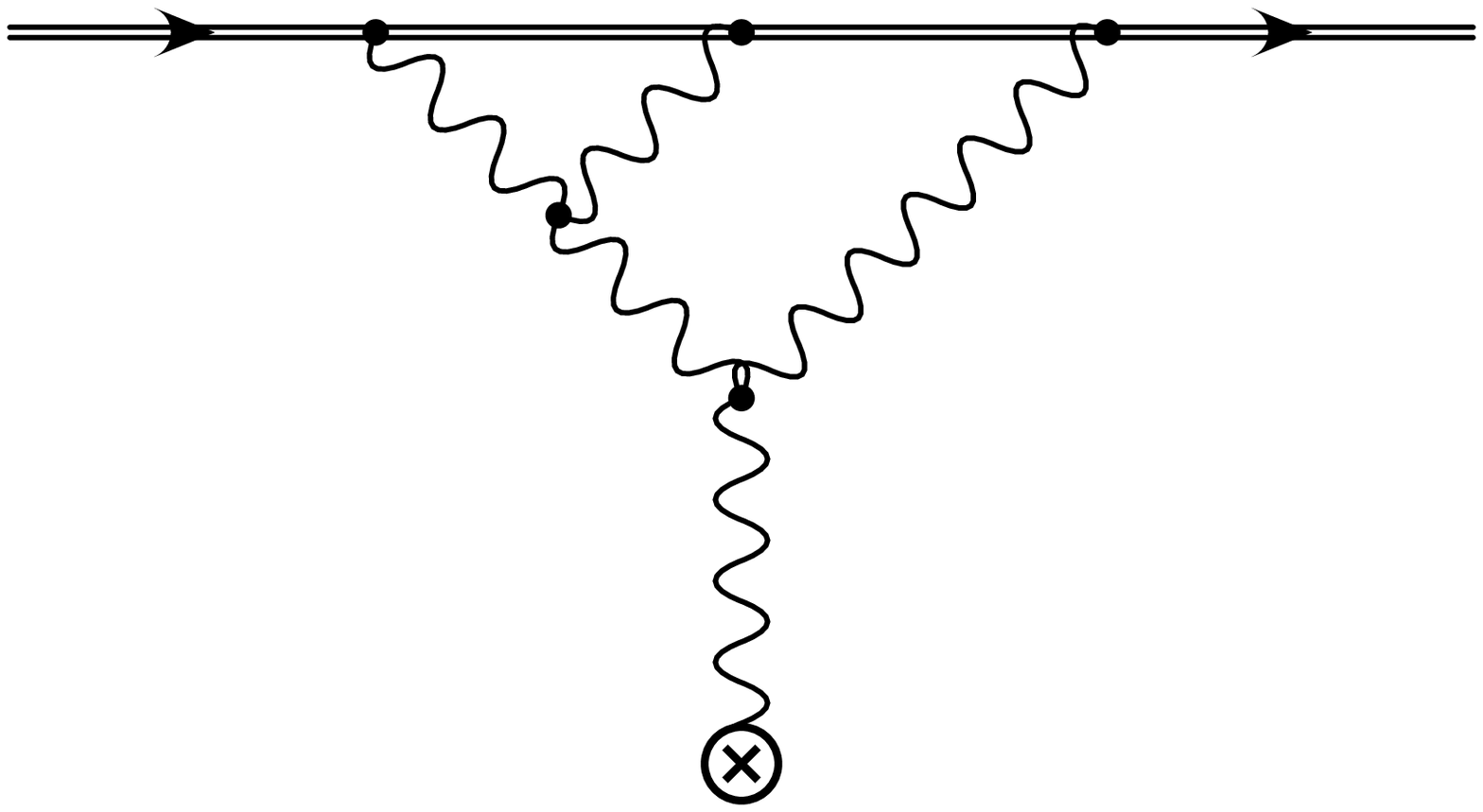} \hfil}\hbox to \size {\hfil(d)\hfil}}}
\vspace{.5cm}
\hbox{\vbox{\hbox to \size {\hfil \includegraphics[width=7cm]{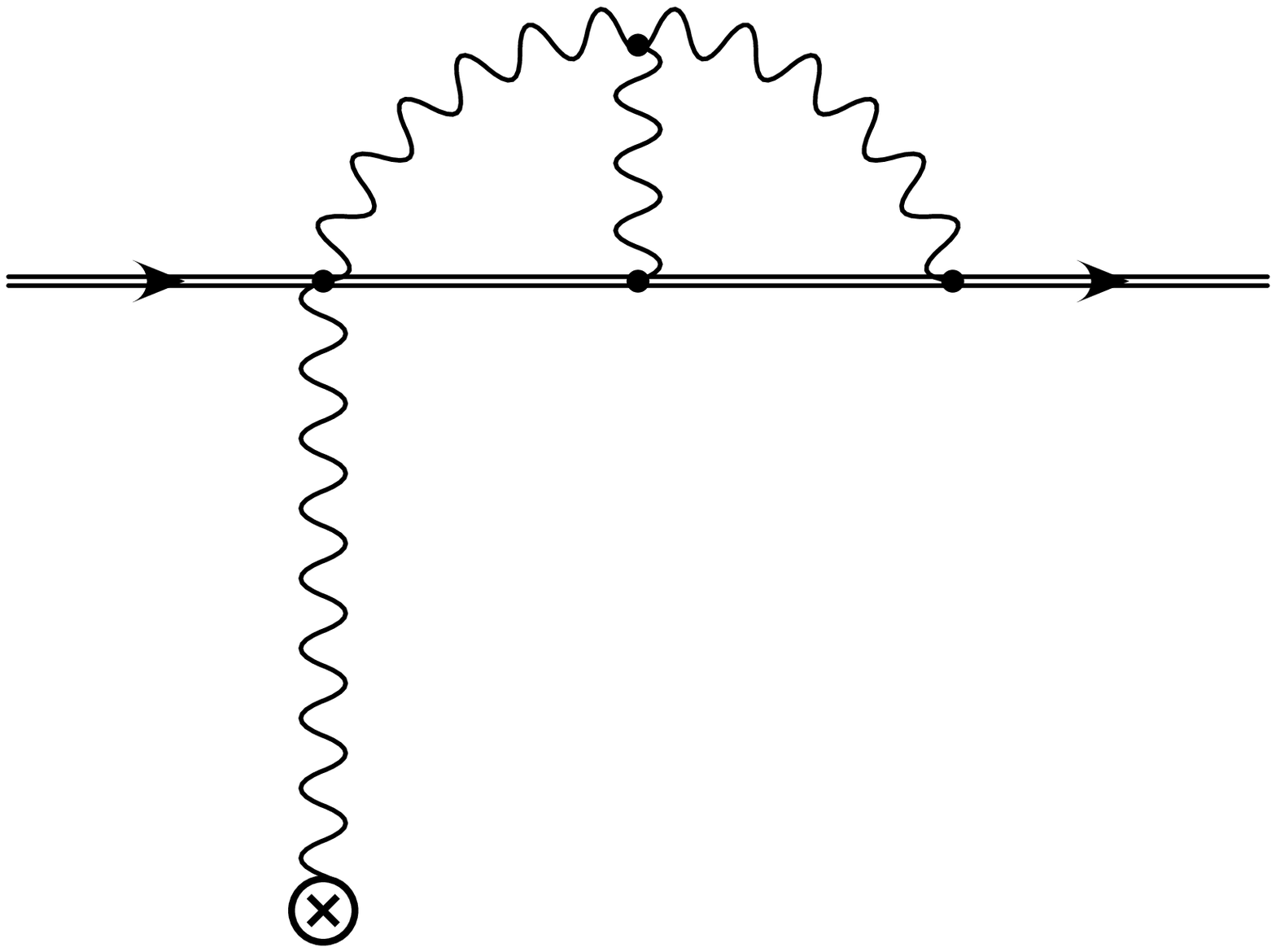} \hfil }\hbox to \size {\hfil(e)\hfil}}
\vbox{\hbox to \size {\hfil \includegraphics[width=7cm]{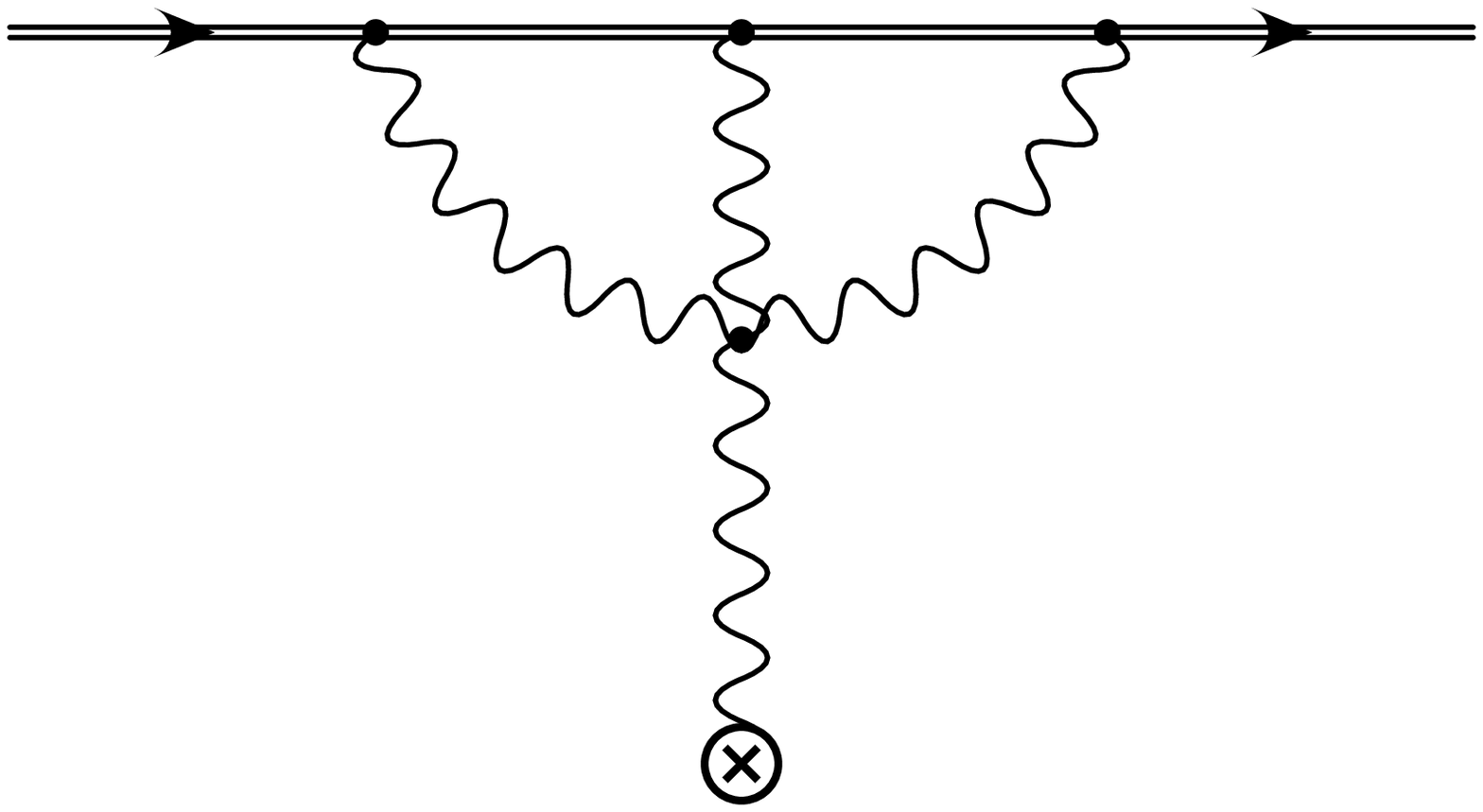} \hfil}\hbox to \size {\hfil(f)\hfil}}}
\caption{Diagrams contributing to the background field tadpoles at order $\ell^{-1}\lambda^3$.  The $\otimes$ denotes an insertion of the background graviton field.}
\label{m3}
\end{figure*}

It is convenient to consider separately the corrections to the $\int dt {\bar h}_{00}$,  $\int dt {\bar h}_{zz}$ terms in the effective action.   For the  $\int dt {\bar h}_{zz}$ terms, the diagrams in Fig.~\ref{m3}(a), Fig.~\ref{m3}(c), and Fig.~\ref{m3}(e) do not contribute.   

It is straightforward to derive the Feynman rules necessary to calculate the diagrams of Fig.~\ref{m3}.   We will simply write down the results of evaluating each diagram.   For the $\int dt {\bar h}_{00}$ tadpoles in the effective action, one finds that evaluating the diagrams at zero external momentum gives rise to no new integrals:   the integration over the internal momenta factorizes into the square of the integrals $I_{0,1}(L)$ of the previous section.   The results are
\begin{eqnarray}
\mbox{Fig.}~\ref{m3}(a) &=& -{i\over 2}m_0 {\hat\lambda}^2 \int dt {\bar h}_{00} ,\\
\mbox{Fig.}~\ref{m3}(b) &=&  {3i\over 2}m_0{\hat\lambda}^2\int dt {\bar h}_{00},\\
\mbox{Fig.}~\ref{m3}(c) &=&  -{3i\over 4}m_0{\hat\lambda}^2\int dt {\bar h}_{00},\\
\mbox{Fig.}~\ref{m3}(d) &=&  -{i\over 4} {13 d- 43\over d-3}m_0 {\hat\lambda}^2\int dt {\bar h}_{00},\\
\mbox{Fig.}~\ref{m3}(e) &=&  {3i\over 2} m_0{\hat\lambda}^2\int dt {\bar h}_{00},\\
\mbox{Fig.}~\ref{m3}(f) &=& {i\over 4}{5 d- 19\over d-3}m_0 {\hat\lambda}^2\int dt {\bar h}_{00}.
\end{eqnarray}
For the tadpole terms $\int dt {\bar h}_{zz}$, we find from Fig.~\ref{m3}(b)
\begin{equation}
\mbox{Fig.}~\ref{m3}(b) = {i\over 2}(d-3)m_0{\hat\lambda}^2\int dt {\bar h}_{zz} .
\end{equation}
The contribution of  graphs Fig.~\ref{m3}(d), Fig.~\ref{m3}(f) to the tadpole $\int dt {\bar h}_{00}$ does not factorize into the integrals of the form $I_{0,1}(L)$.  However their sum does,
\begin{equation}
\mbox{Fig.}~\ref{m3}(d)+\mbox{Fig.}~\ref{m3}(f) = -{i}m_0(d-3) {\hat\lambda}^2\int dt {\bar h}_{zz}.
\end{equation}
Thus the ${\cal O}(\ell\lambda^2)$ terms in the effective action are
\begin{equation}
\left.\Gamma[{\bar h}]\right|_{\ell\lambda^2} = -{1\over 4}m_0{\hat\lambda}^2 \int dt {\bar h}_{00} -{1\over 2} m_0 (d-3) {\hat\lambda}^2 \int dt {\bar h}_{zz}.
\end{equation}

\section{Thermodynamics}
\label{sec:thermo}
We have found, from the diagrams in Fig.~\ref{m2} and Fig.~\ref{m3}
\begin{eqnarray}
{m\over m_0} &=& 1-{1\over 2}{\hat \lambda}+{1\over 2}{\hat\lambda}^2\\
{\tau L\over m_0} &=& {1\over 2} (d-3){\hat \lambda} -  (d-3){\hat\lambda}^2
\end{eqnarray}
To obtain observables which can be tested against the numerical data of~\cite{kolnumeric,kudoh1,kudoh2}, we must eliminate the unphysical bare mass parameter $m_0$ from these two equations.   This gives,
\begin{equation}
\label{eq:tvm}
{2\over d-3}{\tau L\over m} = \zeta(d-3){2 G_N m\over L^{d-3}} - \zeta^2(d-3) \left({2 G_N m\over L^{d-3}}\right)^2+\cdots,
\end{equation}
which agrees with the results of~\cite{5d} when $d=5$.

We may then relate the asymptotic charge to the thermodynamics quantities via
 Smarr's relation $(d-3) m = (d-2) TS +\tau L$ (see ref.~\cite{ho2,koltheory}) which, using Eq.~(\ref{eq:tvm}), gives $TS$ as a function of $m,L$.   As $L\rightarrow\infty$ the entropy $S$ simply becomes the entropy of an isolated $d$-dimensional black hole.   This scales like the area of the black hole, $S\sim r_s^{d-2}\sim m^{{d-2\over d-3}}$.  Thus for $\lambda\ll 1$ we expect
\begin{equation}
S = m^{{d-2\over d-3}} f\left(G_N m\over L^{d-3}\right).
\end{equation}
The function $f(z)$ can be obtained from the relation
\begin{equation}
{1\over T} = \left.{\partial S\over\partial M}\right|_L
\end{equation} 
together with the formula for $TS$ that follows from Smarr's law.  We finds
\begin{equation}
{S\over S(L\rightarrow\infty)} = 1 +{1\over 2} \left({d-2\over d-3}\right) \zeta(d-3) {2 G_N m\over L^{d-3}} + {1\over 8} {d-2\over (d-3)^2} \zeta^2(d-3) \left({2 G_N m\over L^{d-3}}\right)^2 +\cdots,
\end{equation}
and 
\begin{equation}
{T\over T(L\rightarrow\infty)} = 1 - {2d-5\over 2(d-3)}\zeta(d-3) {2 G_N m\over L^{d-3}} +\left[8 d^2 - 43 d + 58\over 8(d-3)^2\right]\zeta^2(d-3) \left({2 G_N m\over L^{d-3}}\right)^2+\cdots,
\end{equation}
where $S(L\rightarrow\infty)$ and $T(L\rightarrow\infty)$ are the entropy and temperature of an uncompactified black hole,
\begin{eqnarray}
S(L\rightarrow\infty) &=&  {2\pi\over (d-2) G_N} r^{d-2}_s, \\
T(L\rightarrow\infty) &=& {d-3\over 4\pi r_s} .
\end{eqnarray}

\begin{figure}[t!]
\def\size{8 cm}
\hbox{\vbox{\hbox to \size {\hfil \includegraphics[width=7cm]{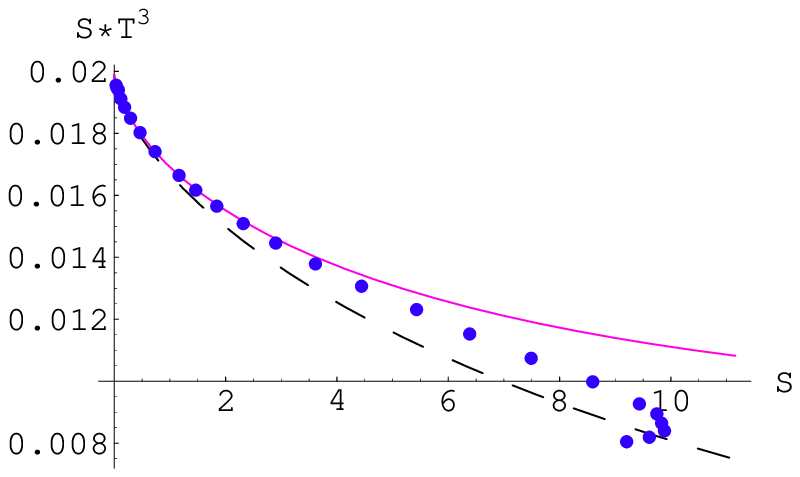} \hfil }\hbox to \size {\hfil(a)\hfil}}
\vbox{\hbox to \size {\hfil \includegraphics[width=7cm]{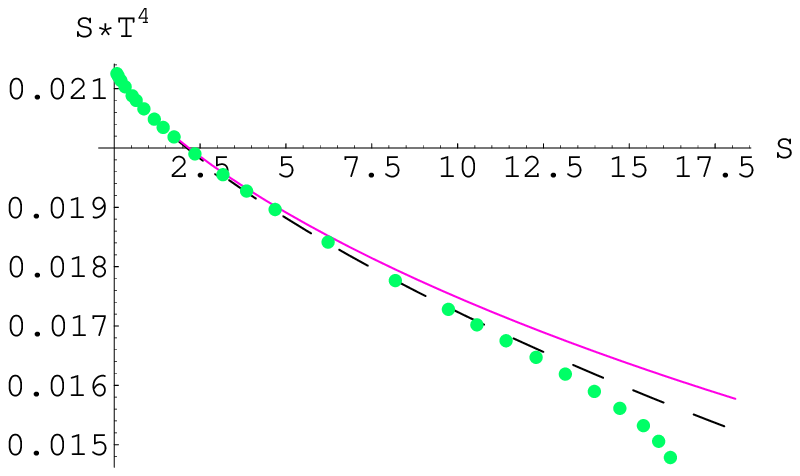} \hfil}\hbox to \size {\hfil(b)\hfil}}}
\caption{Comparison of analytic results with the data of refs.~\cite{kudoh1,kudoh2}.   (a) compares the $d=5$ results and (b) the $d=6$ results.   The dashed lines include only ${\cal O}(\lambda)$ corrections.   The solid lines include ${\cal O}(\lambda^2)$ terms.}
\label{fig:data}
\end{figure}

\begin{figure}[t!]
 \includegraphics[width=8cm]{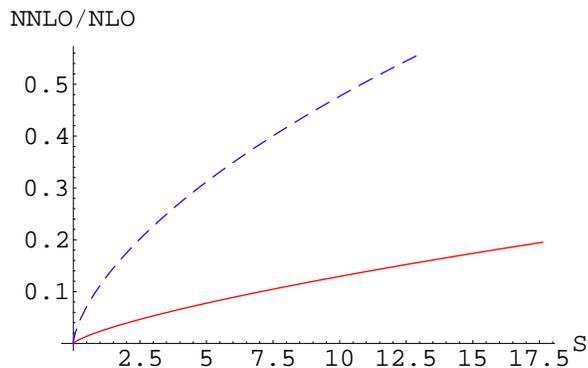} 
\caption{Ratio of ${\cal O}(\lambda^2)$ to ${\cal O}(\lambda)$ terms in the perturbative expansion of $TS^{d-2}$ versus $S$ for $d=5$ (dashed line) and $d=6$ (solid line).}
\label{fig:conv}
\end{figure}

We may then compare with the numerical results of Kudoh and Wiseman \cite{kudoh1,kudoh2} in 
the special cases of five and six dimensions which are shown in Fig.~\ref{fig:data}(a) and Fig.~\ref{fig:data}(b) respectively\footnote{To compare with the numerics, we set $L=\pi$ in units where the entropy of the uncompactified $d$-dimensional black hole is $S=A/4$.}.   The difference between the numerical data and the analytical results grows with $S$, but it is difficult to gauge the relevance of this deviation without some measure of the errors in the numerical computation.   As a crude measure of convergence of the perturbative expansion we plot in Fig.~\ref{fig:conv} the ratio of the ${\cal O}(\lambda^2)$ to the ${\cal O}(\lambda)$ terms in the series expansion for $ST^{d-2}$ versus $S$.    

\section{Conclusions}

In this paper, we have used EFT methods to determine the qualitative structure of the thermodynamics of Kaluza-Klein black holes when their radius is much smaller than the compactification scale.   Using the power counting in the EFT, we find that the asymptotic charges $m,\tau$ are related in the regime $v^2= G_N m/L^{d-2} \ll 1$ by an expansion of the form
\begin{equation}
{\tau L\over m} = f_0(v^2)+ \sum^\infty_{n=1} v^{2\gamma_n} f_n(v^2),
\end{equation}
where $f_n(v^2)$ is analytic about zero and $\gamma_n=2+4n/(d-3)$.  For $n=0$, we find
\begin{equation}
f_0(v^2) = {1\over 2}(d-3)\left[\zeta(d-3) v^2  - \zeta(d-3)^2 v^4 + {\cal O}(v^6)\right],
\end{equation}
which agrees with the ${\cal O}(v^2)$ results of~\cite{kold} in $d$ spacetime dimensions and with the ${\cal O}(v^4)$ $d=5$ results of~\cite{5d} calculated in perturbation theory about the full uncompactifed Schwarzschild background.    Thus our results indicate that the existing analytical tests of the numerics for $d=5,6$ only probe the thermodynamics of point particles and are not sensitive to the dynamics of the black hole horizon.   It would be interesting to repeat the numerical simulations for large dimension where the phase diagram is more sensitive to the structure of the horizon.   For instance in $d\geq 8$, the first finite size effect comes in at order $\lambda^p$, $2<p<3$, and is distinguishable from terms in the black hole thermodynamics that can be reproduced by a minimal point particle action $-m\int ds$.
\\
\\
\centerline{\bf Acknowledgments}

We thank H. Kudoh and T. Wiseman for making the numerical results of~\cite{kudoh1,kudoh2} available to us, and T. Wiseman for helpful discussions.   Y-ZC and WG are supported in part by the Department of Energy under Grant DE-FG02-92ER40704.   IZR is supported in part by the Department of Energy under Grants DOE-ER-40682-143 and DEAC02-6CH03000.

\end{document}